# Modification of structural disorder by hydrostatic-pressure in superconducting YBa$_2$Cu$_3$O$_{6.73}$ cuprate

H. Huang,[1,2,3] H. Jang,[1,†,*] M. Fujita,[4] T. Nishizaki,[5] Y. Lin,[6] J. Wang,[2,3] J. Ying,[7] J. S. Smith,[7] C. Kenney-Benson,[7] G. Shen,[7] W. Mao,[6,8] C.-C. Kao,[9] Y.-J. Liu,[1,*] and J.-S. Lee[1]

[1]*Stanford Synchrotron Radiation Lightsource, SLAC National Accelerator Laboratory, Menlo Park, California 94025, USA*

[2]*State Key Laboratory of Infrared Physics, Shanghai Institute of Technical Physics, Chinese Academy of Sciences, Shanghai 200083, China*

[3]*University of Chinese Academy of Sciences, Beijing 100049, China*

[4]*Institute for Materials Research, Tohoku University, Katahira 2-1-1, Sendai, 980-8577, Japan*

[5]*Department of Electrical Engineering, Kyushu Sangyo University, Fukuoka 813-8503, Japan*

[6]*Stanford Institute for Materials and Energy Sciences, SLAC National Accelerator Laboratory, Menlo Park, California 94025, USA*

[7]*High Pressure Collaborative Access Team, Geophysical Laboratory, Carnegie Institution of Washington, Argonne, Illinois 60439, USA*

[8]*Geological Sciences, Stanford University, Stanford, California 94305, USA*

[9]*SLAC National Accelerator Laboratory, Menlo Park, California 94025, USA*

[†]Present address: PAL-XFEL Beamline Division, Pohang Accelerator Laboratory, 80 Jigokro-127-beongil, Nam-gu, Pohang, Gyeongbuk 37673, Republic of Korea.

[*]Correspondence to: h.jang@postech.ac.kr and liuyijin@slac.stanford.edu

## ABSTRACT

Compelling efforts to improve the critical temperature ($T_c$) of superconductors have been made through high-pressure application. Understanding the underlying mechanism behind such improvements is critically important, however, much remains unclear. Here we studied ortho-III $YBa_2Cu_3O_{6.73}$ (YBCO) using x-ray scattering under hydrostatic-pressure (HP) up to ~6.0 GPa. We found the reinforced oxygen order (OO) of YBCO under HP, revealing an oxygen rearrangement in the Cu-O layer, which evidently shows the charge transfer phenomenon between the $CuO_2$ plane and Cu-O layer. Concurrently, we also observed no disorder-pinned charge density wave (CDW) signature in $CuO_2$ plane under HP. This indicates that the oxygen rearrangement modifies the quenched disorder state in the $CuO_2$ plane. Using these results, we appropriately explain why pressure-condition can achieve higher $T_c$ compared with the optimal $T_c$ under ambient pressure in $YBa_2Cu_3O_{6+x}$. As an implication of these results, finally, we have discussed that the change in disorder could make it easier for $YBa_2Cu_3O_{6+x}$ to undergo a transition to the nematic order under an external magnetic field.

# I. INTRODUCTION

Since high-temperature superconductivity (HTSC) in La-based cuprate was discovered in 1986,[1] cuprates have been a significant focus on HTSC research activities aiming for further improvement in $T_c$.[2-5] The highest reported $T_c$ at ambient pressure is 134 K in Hg-based cuprate.[6] Meantime, the current record-setting $T_c$ at 203 K in the sulfur-hydride system was recently reported under high-pressure application.[7] Indeed, pressure-induced $T_c$ changes have also been reported in many cuprates.[8-12] For example, the Hg-based cuprate, $HgBa_2Ca_2Cu_3O_{8+\delta}$, shows $T_c$ increase up to 164 K at 31 GPa.[9] In $YBa_2Cu_4O_8$ ($T_c$ = 80 K at ambient pressure), $T_c$ increases up to 108 K at 10 GPa.[10] Therefore, concurrent with efforts that survey a large number of materials in varying compositions, applying high-pressure to the superconducting materials has emerged as a promising approach that can efficiently improve superconductivity. Nevertheless, '*why does $T_c$ change upon compression?*' is an essential and fundamental question that still needs to be addressed.

Archetypal Y-based cuprate, $YBa_2Cu_3O_{6+x}$, has been the active subject of many high-pressure HTSC investigations in both experiments[11-18] and theories,[14, 19-23] aiming to address this question. This is in part due to $YBa_2Cu_3O_{6+x}$ being widely appreciated as one of the cleanest cuprates.[24-26] Figure 1(a) represents the $T_c$ behavior for $YBa_2Cu_3O_{6+x}$ without and with high-pressure (P = 0 and P > 0, respectively).[11-17, 27] The $T_c$ is increased up to the optimal doping (where $p \sim 0.165$, $x \sim 0.91$) upon compression. Above the optimal doping range, compression reduces the $T_c$. According to the previous reports,[13, 14, 27] such $T_c$ change has been mainly attributed to the change in bonding lengths along the *c*-axis, in particular, the distance between the Cu-O chain layer and $CuO_2$ plane [Fig. 1(b)]. Compression brings the Cu-O chain layer closer to the $CuO_2$ plane, which enhances charge transfer between the planes, leading to additional hole-doping in the $CuO_2$ plane, as theoretically explained.[14, 20-23] In addition, the in-plane oxygen movement in the structurally disordered Cu-O chain layer is considered as an additional hole-doping source into the $CuO_2$ plane. Simultaneously, such compression induced oxygen movement modifies the oxygen order (OO) in YBCO.[14-16, 18, 23,28] However, this hypothesis is not fully established yet, due to the absence of direct experimental evidence. Moreover, the reported maximum $T_c$ under high-pressure ($T_c^{max}|_{P>0} \sim 107$ K in $YBa_2Cu_3O_{6.66}$ and $p \sim 0.134$)[12, 17] is considerably higher than that under ambient pressure ($T_c^{max}|_{P=0} \sim 94$ K).[27] This indicates that charge transfer mechanism (i.e., doping change) is unlikely the sole reason for the pressure-induced $T_c$ improvement. This is because, if the change in $T_c$ upon compression is only driven by the doping, the $T_c^{max}|_{P>0}$ must be the same with or, at least, similar to the $T_c^{max}|_{P=0}$. In this context, a charge density wave (CDW), which is competing with YBCO's superconductivity[29-39] and understood as the reason of the $T_c$ plateau around $p \sim 1/8$,[29-31,34,39] has been recently proposed as an additional driving force for the $T_c$ change under high-pressure.[17] However, there is a lack of evidence to prove this proposal. As a result, we are still in early stages of understanding the underlying mechanism(s) for the pressure-induced $T_c$ changes.

In this work, we performed x-ray scattering measurements on an under-doped $YBa_2Cu_3O_{6.73}$ (YBCO) ortho-III crystal ($p \sim 0.13$, $T_c \sim 70$ K). We observed that the pressure modifies the disorder state in YBCO, leading to an enhancement of OO in the Cu-O layer and a suppression of disorder-pinned CDW fluctuations in the $CuO_2$ plane. Through these results, we could verify the role of the structural disorder. Finally, we understand the reason why pressure can increases the $T_c$ in YBa$_2$Cu$_3$O$_{6+x}$ to a level that is considerably higher than the optimal $T_c$ under ambient pressure.

## II. EXPERIMENTAL DETAILS

High-quality single crystals of YBa$_2$Cu$_3$O$_{6.73}$ were grown by self-flux method using Y$_2$O$_3$ crucible. The oxygen content ($x$) was controlled by the annealing condition under oxygen flow atmosphere. The superconducting transition temperature $T_c$ was determined by the magnetization measurements using SQUID magnetometer (MPMS3). The measured YBCO crystals annealed at 630 °C for 7 days show sharp superconducting transitions with $T_c \sim 70$ K. Since this single crystal prepared by the same method shows the first-order vortex lattice melting transition, the single crystals studied in this work are clean and homogeneous single crystals, in spite of no detwinning-process.

For this study, we employed two scattering approaches; resonant soft x-ray scattering (RSXS) and high-pressure hard x-ray diffraction. First, the RSXS experiments were performed at beamline 13-3 of the Stanford Synchrotron Radiation Lightsource (SSRL). The *ex situ* cleaved YBCO crystal was mounted on an in-vacuum 4-circle diffractometer and cooled down with an open-flow liquid helium cryostat. ($h$, 0, $l$) scattering plane was explored by rotating sample angle ($\theta$) after aligning the crystalline ac-plane parallel to the scattering plane. Two-dimensional (2D) CCD detector was utilized to detect a diffraction pattern over a wide reciprocal space. The center of CCD was fixed at the scattering angle, $2\theta = 156°$. Because of quasi two-dimensional nature of the OO and CDW,[34, 40] the peaks are observed in broad $l$ range, and the value ($l = \sim 1.4$ in current experiment) is determined by the $2\theta$ position. The photon polarization was perpendicular to the scattering plane (σ-polarization), which ensures strong signal for both the oxygen order (OO) and charge density wave (CDW)[31]. Second, the high-pressure hard x-ray diffraction experiments were performed at beamlines 16-ID-B and 16-BM-D of the Advanced Photon Source (APS). The YBCO single crystal with dimensions of $a \times b \times c \approx 80 \times 60 \times 40 \ \mu m^3$ was mounted in the diamond anvil cell (DAC) with compressed helium gas loaded as the pressure-medium to ensure the hydrostatic-pressure (HP) environment up to our maximum pressure. The DAC system was also installed on the open-circle cryostat, aiming for detecting a CDW signal around $T_c$. The 30 keV, high-energy x-ray, and a large 2D area detector were utilized to ensure that the $hk$-diffraction pattern could capture many of the lattice Bragg peaks, as well as the OO patterns.

## III. X-RAY SCATTERING RESULTS AND DISCUSSION

In $YBa_2Cu_3O_{6+x}$, structural properties such as OO and CDW have been widely appreciated as important ingredients for understanding its superconductivity. As a first step, we investigated these properties at ambient pressure in YBCO using the Cu $L_3$-edge RSXS. In the ideal ortho-III structure in YBCO, two fully filled ($F$) Cu-O chains and an oxygen-empty ($E$) Cu-line along the crystalline $b$-axis in the Cu-O chain layer show a periodic formation along the $a$-axis: i.e., $F$-$F$-$E$-$F$-$F$-$E$-$F$... sequence (Fig. 1(b) – upper panel), resulting in OO with the wavevector $q = 1/3$.[31, 40] In this YBCO, we observed an OO peak at $\boldsymbol{Q}_{OO}$ = (-0.33, 0, $l$) under 300 K and P = 0 GPa [Fig. 1(c)]. Since CDW is also considered as an additional driving mechanism for understanding the $T_c$ improvement under compression,[17] we investigated CDW order in this YBCO. Figure 1(d) shows the short ranged CDW order at $Q_{CDW}$ = (-0.31, 0, $l$) at $T_c$ = 70 K. These scattering observations are consistent with previous reports.[31, 34] Note that the same CDW order in this doped YBCO was reported in hard x-ray scattering study under ambient pressure.[32]

To explore the change in YBCO properties under high-pressure, hard x-ray diffraction measurement with 30 keV photon energy was performed on the same crystal. Figure 2(a) shows the scattering geometry (transmission-configuration). As the volume of crystal unit-cell shrinks under compression generally, we infer accordingly that the positions of the lattice Bragg peaks move to larger 2$\theta$ angle. As shown in Fig. 2(b), both 2$\theta$ angles of YBCO's in-plane Bragg peaks, (2, 0, 0) and (0, 2, 0) move toward the larger angles. Interestingly, the pressure-induced 2$\theta$ shifts are different to these (2, 0, 0) and (0, 2, 0) reflections. With increasing the HP, the shift of (2, 0, 0) is larger than that of (0, 2, 0) [Fig. 2(c), upper-panel]. This indicates that the reduction of the lattice parameter along the $a$-axis is greater than that along the $b$-axis, increasing the orthorhombicity ($b/a$) of YBCO even with HP (i.e., isotropic pressure) [Fig. 2(c), lower-panel].

Under HP condition, the change in orthorhombicity brings our attention to exploring the OO of YBCO. This is because, as shown in Fig. 1(b), the orthorhombic structure of YBCO is directly correlated with the Cu-O chain layer that has the intrinsic anisotropy (chain) feature. For this reason, we explored the OO in YBCO to investigate – whether it could be related to the increase in orthorhombicity under HP. Figure 3(a) shows the OO patterns at $Q_{OO}$ = (3-$q$, 0, $l$) at 1.0 and 5.8 GPa at room temperature. Note that the entire patterns are shown in Fig. 1(a). Due to the intrinsic quasi 2-dimensional feature of OO,[40] the $l$-value is rather insensitive to observe OO. Relatively, the intensity of the OO peak (i.e., peak height) gets stronger at P = 5.8 GPa, while a change of the peak-width (i.e., correlation length) along either the *h* or *k*-direction is not obvious. For better statistics, we analyzed more OO peaks within our detectable windows, which are summarized in Fig. 3(b). Under HP condition, we could find the clear enhancement of OO peaks' intensity compared with the zero pressure level. Note that the absolute values of the peak-height (*y*-axis) have been shifted for comparison together, because of their different geometric effects as well as the different scattering structure-factor related with the corresponding nearest lattice Bragg peaks. From this result, we infer that the compression causes an oxygen re-distribution inside the Cu-O chain layer. Moreover, as shown in the inset of Fig. 3(b), the width change is not clear, also inferring that a changing in domain size under HP is negligible in this experimental resolution.

In order to gain further insight into the changes in the OO under HP, we simulated the OO patterns on the OO in reciprocal space by varying the oxygen distribution in the Cu-O layer (i.e., real-space). For this simulation, we employed 2-dimensional cell by 12 (*a*-axis) × 36 (*b*-axis). Since the width-change is not experimentally obvious in our resolution, we have constrained the fixed domain size (i.e., the finite cell-size) for this simulation. And besides, to make simple ortho-III structure, oxygen atoms filled 2/3 of cell spaces and *F*–*F*–*E*–*F*–… chain patterns of oxygen distributions were generated along the *a*-axis. The number of oxygen atoms in *F* and *E* chains was controlled by '*randomness*' – defined by [(number of oxygen atoms in empty chains) / (2/3 × number of sites in all empty chains)]. For example, 50 % randomness means that 1/3 of cell spaces in the empty chains are filled by the oxygen while the zero-randomness indicates no-oxygen in the empty chains – i.e., the ideal ortho-III case [Fig. 4(a)]. The fact that this *randomness* is how much different from the ideal (i.e., perfect case) ortho-III structure in the Cu-O layer.

Figure 4(b) shows the simulated OO diffraction patterns on two cases – 50 % randomness (left-panel) and 25 % (right). As reducing the randomness (i.e., oxygen's locations are at the proper place), the pattern becomes strong and clear. Moreover, with varying the randomness, we could systematically trace the height of OO peak [Fig. 4(c)]. Note that aiming for avoiding unrealistic cases (e.g., no-OO (i.e., 100 % randomness) or ideal ortho-III structure), the OO patterns between 75 % and 25 % randomness have been tested. In this simulation, overall OO behaviors resemble the experimental observations in the OO ones under HP. This analogy between the experiment and the simulation indicates that the randomness of oxygen distribution in the Cu-O chain layer is decreasing under HP. In this context, the reduced randomness by HP improves OO, leading to the anisotropic stacking force in Cu-O layer (i.e., a bonding length changes[42]), which would increase the orthorhombicity.[43, 44]

Furthermore, the oxygen rearrangement in the Cu-O chain under HP affects the charge configuration in the Cu-O chains, as well as a distance between Cu-O layer and $CuO_2$ plane.[13] Concurrently, the electronic structure (e.g., valence band) of Cu atoms can be varying, depending on whether Cu is adequately bonded with O or not.[45, 46] Figure 5 shows a schematic diagram of Cu states' modification. While the oxygen rearrangement happens in the Cu-O chain, additional electrons transfer from the $CuO_2$ plane into the Cu-O chain layer to maintain the charge neutrality. Eventually, this HP effect causes an extra hole doping effect in the $CuO_2$ plane, leading to an increase of $T_c$ of underdoped $YBa_2Cu_3O_{6+x}$.

To understand the $T_c^{max}|_{P>0}$ behavior in YBCO which is higher than $T_c^{max}|_{P=0}$ (see Fig. 1(a), and corresponding the text), however, in fact, this charge transfer cannot be solely a reason. As we discussed in the chapter of introduction, the natural interpretation is that decreased strength of the CDW order turns to increase $T_c$ more. In this sense, we investigated a CDW under HP. Note that this YBCO shows the clear CDW under the ambient pressure [see Fig. 1(d)]. Figure 6(a) shows the diffraction patterns around $\boldsymbol{Q} = (3+q^{CDW}, 0, l)$ under P = 1.0 GPa (top-panel: $T$ = 150 K, middle: 70 K ~ $T_c$). Where, the $q^{CDW}$ is around 0.31 r.l.u. Note that a scattering signal of CDW in $YBa_2Cu_3O_{6+x}$ is maximized around $T_c$.[34] While the oxygen order is clearly seen at both temperatures, no CDW signature has been seen in our detecting resolution. To confirm whether there is no signature or not, we also plot a difference (bottom-panel) between two temperatures. It only shows a tiny residual signature of OO. There are two possibilities (extrinsic and intrinsic) to explain this CDW absence. First, the extrinsic reason is that the generally weak CDW signal is buried by a high background signal from diamond in the DAC. Second, the intrinsic reason can be that the suppressed CDW is due to HP effect. Unfortunately, at this moment, we cannot distinguish them clearly. However, considering the recent inelastic x-ray scattering result,[47] the suppression of CDW at even lower pressure is likely intrinsic. In addition, our OO results support this behavior seems to be intrinsic. Since the fact that the modified structural disorder in the Cu-O layer (i.e., oxygen rearrangement) affects the charge distribution (i.e., charge-density potential) in the $CuO_2$ plane, the quenched disorder in the $CuO_2$ plane, which originates from the charge potential of defects,[37] is correspondingly modified through the pressure. This circumstance may reduce the pinning centers for CDW fluctuation in the $CuO_2$ plane,[36] suppressing the disorder-pinned CDW fluctuation, which is consistent with the theoretical predictions.[33, 37, 38] As a consequence, we come to an answer to the initial question we posed. There are two factors contributing to the compression induced $T_c$ change. Firstly, in agreement with the previous works,[14, 20-23] high-pressure promoted the charge transfer (i.e., hole-doping into the $CuO_2$ plane) leads to the increment in $T_c$. Secondly, the quenched disorder state is modified, resulting in the suppression of the CDW. Subsequently, due to the competing nature between YBCO's superconductivity and CDW,[29-39] $T_c^{max}|_{P>0}$, which is even higher than the optimal $T_c$ at ambient pressure, can be achieved through compression.

Finally, we try to extend these results to reconcile the broader context of CDW phenomena in underdoped $YBa_2Cu_3O_{6+x}$ [see Fig. 6(b)]. Considering theoretical predictions[48] and corresponding experimental results[39], the ideal CDW state (i.e., ground state) in YBCO is of long-range, anisotropic (or nematic) ordering when superconductivity is suppressed by the external magnetic field ($H$-field). Theoretically, this nematic order also prefers structural nematicity (i.e., orthorhombicity).[37] By enhancing $YBa_2Cu_3O_{6+x}$'s orthorhombicity under hydrostatic-pressure, it would become easier to develop nematic order in compressed YBCO (state-II, labeled in Fig. 6(b)) compared to ambient pressure YBCO (state-I). On the other hand, the oxygen redistribution in the Cu-O layer under high-pressure suppresses the disorder-pinned CDW fluctuating portion. Therefore, we infer that a pressure condition causes to decrease the disorder strength ($\sigma$)[37,39,48] in the $CuO_2$ plane, which is reasonably proportional to or dominated by the quantity of the structural disorders in Cu-O chain layer.[36,49] Ultimately, the energy cost for developing the nematic order at the state-II ($E_{II}$, i.e. $H$-field strength) can be smaller than that at the state-I ($E_I$),[39] which is well supported by our findings (i.e., $\sigma_{II} < \sigma_I$ and $\Lambda_{II} < \Lambda_I$).

## IV. SUMMARY

In summary, we present the detailed study of structural disorder in YBCO using x-ray scattering under high-pressure. We observed the enhanced oxygen order under hydrostatic-pressure, resulting in the broken of quenched disorder state and the decreased disorder which leads to changes in the hole-doping as well as the weakening of the CDW fluctuation. These findings suggest the underlying mechanism of the pressure-induced $T_c$ change in YBCO. Simultaneously, the pressure induced clean-nature (i.e. suppressed disorder) could be favorable to the development of nematic order in YBCO. This potential development suggests that there is a pressure threshold, above which further compression has limited influence on $T_c$.

# REFERENCES

1.  J. G. Bednorz, K. A. Müller, Possible high $T_c$ superconductivity in the Ba−La−Cu−O system. *Z. Phys. B* **64**, 189-193 (1986).

2.  M. K. Wu *et al.* Superconductivity at 93 K in a new mixed-phase Y-Ba-Cu-O compound system at ambient pressure. *Phys. Rev. Lett.* **58**, 908 (1987).

3.  Z. Z. Sheng, A. M. Hermann, Superconductivity in the rare-earth-free Tl–Ba–Cu–O system above liquid-nitrogen temperature. *Nature (London)* **332**, 55-58 (1988).

4.  P. A. Lee, N. Nagaosa, X. G. Wen, Doping a Mott insulator: Physics of high-temperature superconductivity. *Rev. Mod. Phys.* **78**, 17 (2006).

5.  B. Keimer, S. A. Kivelson, , M. R. Norman, , S. Uchida, J. Zaanen,  From quantum matter to high-temperature superconductivity in copper oxides. *Nature (London)* **518**, 179-186 (2015).

6.  A. Schilling,  M. Cantoni,  J. D. Guo,  H. R. Ott, Superconductivity above 130 K in the Hg–Ba–Ca–Cu–O system. *Nature (London)* **363**, 56-58 (1993).

7.  A. P. Drozdov,  M. I. Eremets, I. A. Troyan, V.  Ksenofontov,  S. I. Shylin, Conventional superconductivity at 203 kelvin at high pressures in the sulfur hydride system. *Nature (London)* **525**, 73-76 (2015).

8.  C. W. Chu, *et al.* Evidence for superconductivity above 40 K in the La-Ba-Cu-O compound system. *Phys. Rev. Lett.* **58**, 405 (1987).

9.  C. W. Chu, *et al.* Superconductivity above 150 K in HgBa2Ca2Cu3O8+ δ at high pressures. *Nature (London)* **365**, 323-325 (1993).

10. S. Souliou, *et al.* Pressure-induced phase transition and superconductivity in $YBa_2Cu_4O_8$. *Phys. Rev. B* **90**, 140501 (2014).

11. P. L. Alireza, *et al*. Accessing the entire overdoped regime in pristine $YBa_2Cu_3O_{6+x}$ by application of pressure. *Phys. Rev. B* **95** 100505 (2017).

12. S. Sadewasser, J. S. Schilling, A. P. Paulikas, B. W. Veal, Pressure dependence of $T_c$ to 17 GPa with and without relaxation effects in superconducting $YBa_2Cu_3O_x$. *Phys. Rev. B* **61**, 741 (2000).

13. J. D. Jorgensen, *et al*. Pressure-induced charge transfer and d$T_c$/dP in $YBa_2Cu_3O_{7-x}$. *Physica C* **171**, 93-102 (1990).

14. C. C. Almasan, *et al*. Pressure dependence of $T_c$ and charge transfer in YBa2Cu3Ox (6.35≤ x≤ 7) single crystals. *Phys. Rev. Lett.* **69**, 680 (1992).

15. J. Metzger, *et al*. Separation of the intrinsic pressure effect on $T_c$ of $YBa_2Cu_3O_{6.7}$ from a $T_c$ enhancement caused by pressure-induced oxygen ordering. *Physica C* **214**, 371-376 (1993).

16. S. Sadewasser, *et al*. Pressure-dependent oxygen ordering in strongly underdoped $YBa_2Cu_3O_{7-y}$. *Phys. Rev. B* **56**, 14168 (1997).

17. O. Cyr-Choinière, *et al*. Suppression of charge order by pressure in the cuprate superconductor $YBa_2Cu_3O_y$: Restoring the full superconducting dome. preprint at http://arxiv.org/abs/1503.02033 (2015).

18. T. Tomita, J. S. Schilling, L. Chen, B. W. Veal, H. Claus, Enhancement of the Critical Current Density of $YBa_2Cu_3O_x$ Superconductors under Hydrostatic Pressure. *Phys. Rev. Lett.* **96**, 077001 (2006).

19. W. E. Pickett, Effect of uniaxial strains on the electronic structure of $YBa_2Cu_3O_7$. *Physica C* **289**, 51-62 (1997).

20. J. J. Neumeier, H. A. Zimmermann, Pressure dependence of the superconducting transition temperature of $YBa_2Cu_3O_7$ as a function of carrier concentration: A test for a simple charge-transfer model. *Phys. Rev. B* **47**, 8385 (1993).

21. R. P. Gupta, M. Gupta, Relationship between pressure-induced charge transfer and the superconducting transition temperature in $YBa_2Cu_3O_{7-\delta}$ superconductors. *Phys. Rev. B* **51**, 11760 (1995).

22. X. J. Chen, H. Q. Lin, C. D. Gong, Pressure dependence of $T_c$ in Y-Ba-Cu-O superconductors. *Phys. Rev. Lett.* **85**, 2180 (2000).

23. P. Gao, R. Zhang, X. Wang, Pressure induced self-doping and dependence of critical temperature in stoichiometry $YBa_2Cu_3O_{6.95}$ predicted by first-principle and BVS calculations. *AIP Adv.* **7**, 035215 (2017).

24. H. Alloul, J. Bobroff, M. Gabay, P. J. Hirschfeld, Defects in correlated metals and superconductors. *Rev. Mod. Phys.* **81**, 45 (2009).

25. F. Coneri, S. Sanna, , K. Zheng , J. Lord, R. De Renzi, Magnetic states of lightly hole-doped cuprates in the clean limit as seen via zero-field muon spin spectroscopy. *Phys. Rev. B* **81**, 104507 (2010).

26. G. Grissonnanche, *et al*. Direct measurement of the upper critical field in cuprate superconductors. *Nat. Commun.* **5**, 3280 (2014).

27. R. Liang, D. A. Bonn, W. N. Hardy, Evaluation of $CuO_2$ plane hole doping in $YBa_2Cu_3O_{6+x}$ single crystals. *Phys. Rev. B* **73**, 180505 (2006).

28. S. I. Schlachter *et al*. The effect of chemical doping and hydrostatic pressure on $T_c$ of $Y_{1-y}Ca_yBa_2Cu_3O_x$ single crystals. *Physica C* **328**, 1-13 (1999).

29. G.Ghiringhelli *et al*. Long-Range Incommensurate Charge Fluctuations in (Y,Nd)Ba$_2$Cu$_3$O$_{6+x}$. *Science* **337**, 821-825 (2012).

30. J. Chang *et al*. Direct observation of competition between superconductivity and charge density wave order in YBa$_2$Cu$_3$O$_{6.67}$. *Nat. Phys.* **8**, 871-876 (2012).

31. A. J. Achkar *et al*. Distinct charge orders in the planes and chains of ortho-III-ordered YBa$_2$Cu$_3$O$_{6+\delta}$ superconductors identified by resonant elastic x-ray scattering. *Phys. Rev. Lett.* **109**, 167001 (2012).

32. E. Blackburn *et al*. X-ray diffraction observations of a charge-density-wave order in superconducting ortho-II YBa$_2$Cu$_3$O$_{6.54}$ single crystals in zero magnetic field. *Phys. Rev. Lett.* **110**, 137004 (2013).

33. L. E. Hayward, D. G. Hawthorn, R. G. Melko, S. Sachdev. Angular fluctuations of a multicomponent order describe the pseudogap of YBa$_2$Cu$_3$O$_{6+x}$. *Science* **343**, 1336-1339 (2014)

34. S. Blanco-Canosa *et al*. Resonant x-ray scattering study of charge-density wave correlations in YBa$_2$Cu$_3$O$_{6+x}$. *Phys. Rev. B* **90**, 054513 (2014).

35. S. Gerber *et al*. Three-dimensional charge density wave order in YBa$_2$Cu$_3$O$_{6.67}$ at high magnetic fields. *Science* **350**, 949-952 (2015).

36. T. Wu *et al*. Incipient charge order observed by NMR in the normal state of YBa$_2$Cu$_3$O$_y$. *Nat. Commun.* **6**, 6438 (2015).

37. L. Nie *et al*. Fluctuating orders and quenched randomness in the cuprates. *Phys. Rev. B* **92**, 174505 (2015).

38. Y. Caplan , G. Wachtel, D. Orgad, Long-range order and pinning of charge-density waves in competition with superconductivity. *Phys. Rev. B* **92**, 224504 (2015).

39. H. Jang *et al*. Ideal charge-density-wave order in the high-field state of superconducting YBCO. *Proc. Nat. Acad. Sci.* **113**, 14645-14650 (2016).

40. M. V. Zimmermann *et al*. Oxygen-ordering superstructures in underdoped $YBa_2Cu_3O_{6+x}$ studied by hard x-ray diffraction. *Phys. Rev. B* **68**, 104515 (2003).

41. G. Shen, H. K. Mao, High-pressure studies with x-rays using diamond anvil cells. *Prog. Phys.* **80**, 016101 (2017).

42. E. S. Božin *et al*. Charge-screening role of c-axis atomic displacements in $YBa_2Cu_3O_{6+x}$ and related superconductors. *Phys. Rev. B* **93** 054523 (2016).

43. J. D. Jorgensen *et al*. Time-dependent structural phenomena at room temperature in quenched $YBa_2Cu_3O_{6.41}$: local oxygen ordering and superconductivity. *Physica C* **167**, 571 (1990).

44. B. W. Veal *et al*. Time-dependent superconducting behavior of oxygen-deficient $YBa_2Cu_3O_x$: possible annealing of oxygen vacancies at 300 K. *Phys. Rev. B* **42**, 4770 (1990).

45. V. M. Matic, N. D. Lazarov, Impact of chain fragmentation on charge transfer scenario and two-plateaus-like behavior of $T_c$ $(x)$ in $YBa_2Cu_3O_{6+x}$. *Solid State Commun.* **142**, 165-168 (2007).

46. E. S. Božin *et al*. Charge-screening role of c-axis atomic displacements in $YBa_2Cu_3O_{6+x}$ and related superconductors. *Phys. Rev. B* **93**, 054523 (2016).

47. S. M. Souliou *et al*. Rapid suppression of the charge density wave in $YBa_2Cu_3O_{6.6}$ under hydrostatic pressure. *Phys. Rev. B* **97**, 020503(R) (2018).

48. L. Nie, G. Tarjus, S. A. Kivelson, Quenched disorder and vestigial nematicity in the pseudogap regime of the cuprates. *Proc. Nat. Acad. Sci.* **111**, 7980-7985 (2014).

49. J. S. Bobowski *et al*. Oxygen chain disorder as the weak scattering source in YBa$_2$Cu$_3$O$_{6.50}$, *Phys. Rev. B* **82**, 134526 (2010).

## Acknowledgments

We thank Laimei Nie and Steven A. Kivelson for valuable discussions and comments. Soft x-ray experiments were carried out at the Stanford Synchrotron Radiation Lightsource (SSRL), SLAC National Accelerator Laboratory, is supported by the U.S. Department of Energy, Office of Science, Office of Basic Energy Sciences under Contract No. DE-AC02-76SF00515. W. M., and Y. L. acknowledge support by the Department of Energy, Office of Basic Energy Sciences, Materials Sciences and Engineering Division, under Contract No. DE-AC02-76SF00515. H. H. acknowledges support of China Scholarship Council. J.S.S. C.K.B, and G.S. acknowledge support of DOE-MSED under Award No. DE-FG02-99ER45775. High pressure studies were performed at HPCAT (Sector 16), Advanced Photon Source (APS), Argonne National Laboratory. HPCAT operations are supported by DOE-NNSA under Award No. DE-NA0001974, with partial instrumentation funding by NSF. The Advanced Photon Source is a U.S. Department of Energy (DOE) Office of Science User Facility operated for the DOE Office of Science by Argonne National Laboratory under Contract No. DE-AC02-06CH11357. M.F. and T. N. are supported by Grant-in-Aid for Scientific Research (A) (16H02125) and for Scientific Research (C) (16K05460), respectively.

**FIGURES**

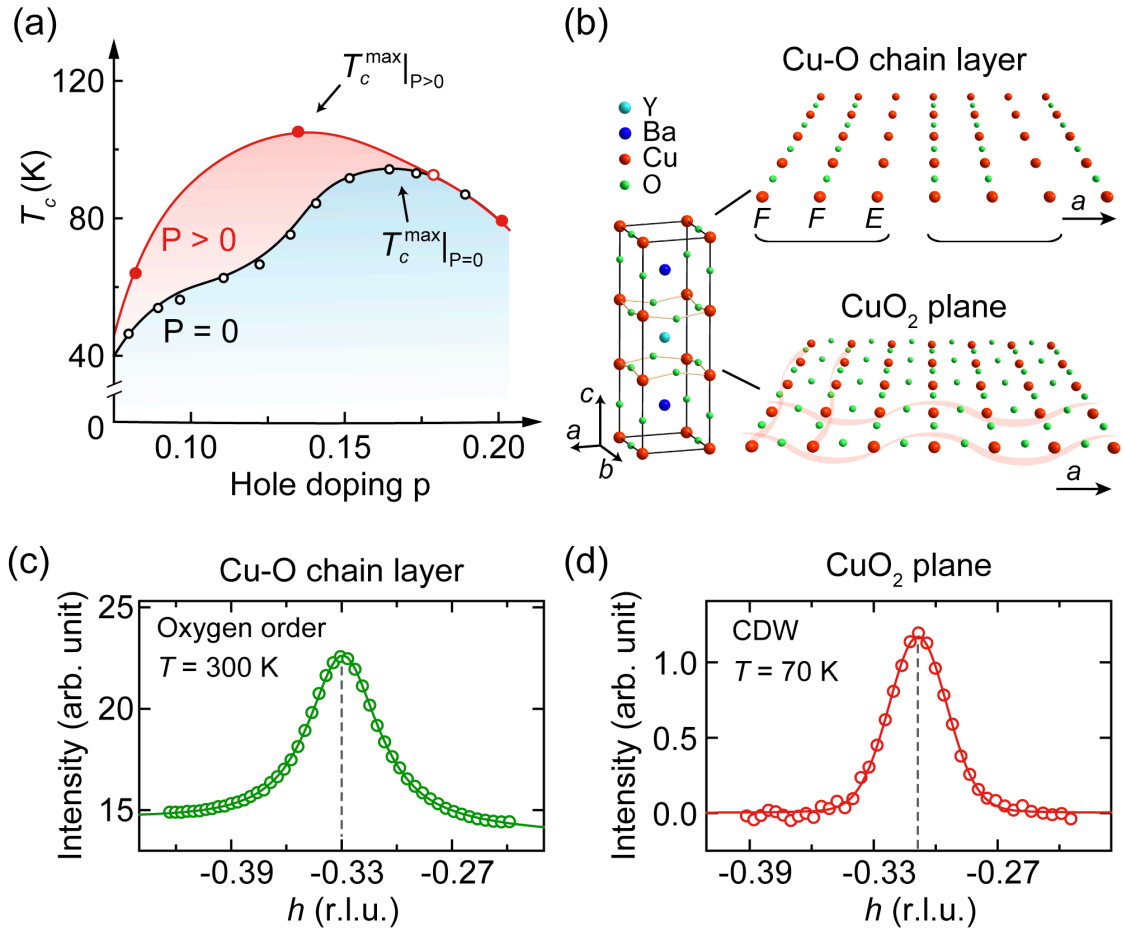

**Fig. 1:** (a) $T_c$ phase diagram of YBCO with/without the high-pressure. The arrows indicate the optimal $T_c$ values. The black circles denote ambient pressure data[27]. Red-colored filled (opened) circles are at 15 GPa (2 GPa)[12,17]. (b) The crystal structure of YBCO (left) and extended drawings of the Cu-O chain layer with OO and CuO$_2$ plane with CDW. (c,d) RSXS data – Ortho-III OO and CDW – measured at 933.7 eV and 930.8 eV, respectively. Dashed lines indicate peak-positions of OO (at $h$ = -0.33 r.l.u.) and CDW (at $h$ = -0.31 r.l.u.). Solid lines are curves fitted to the data.

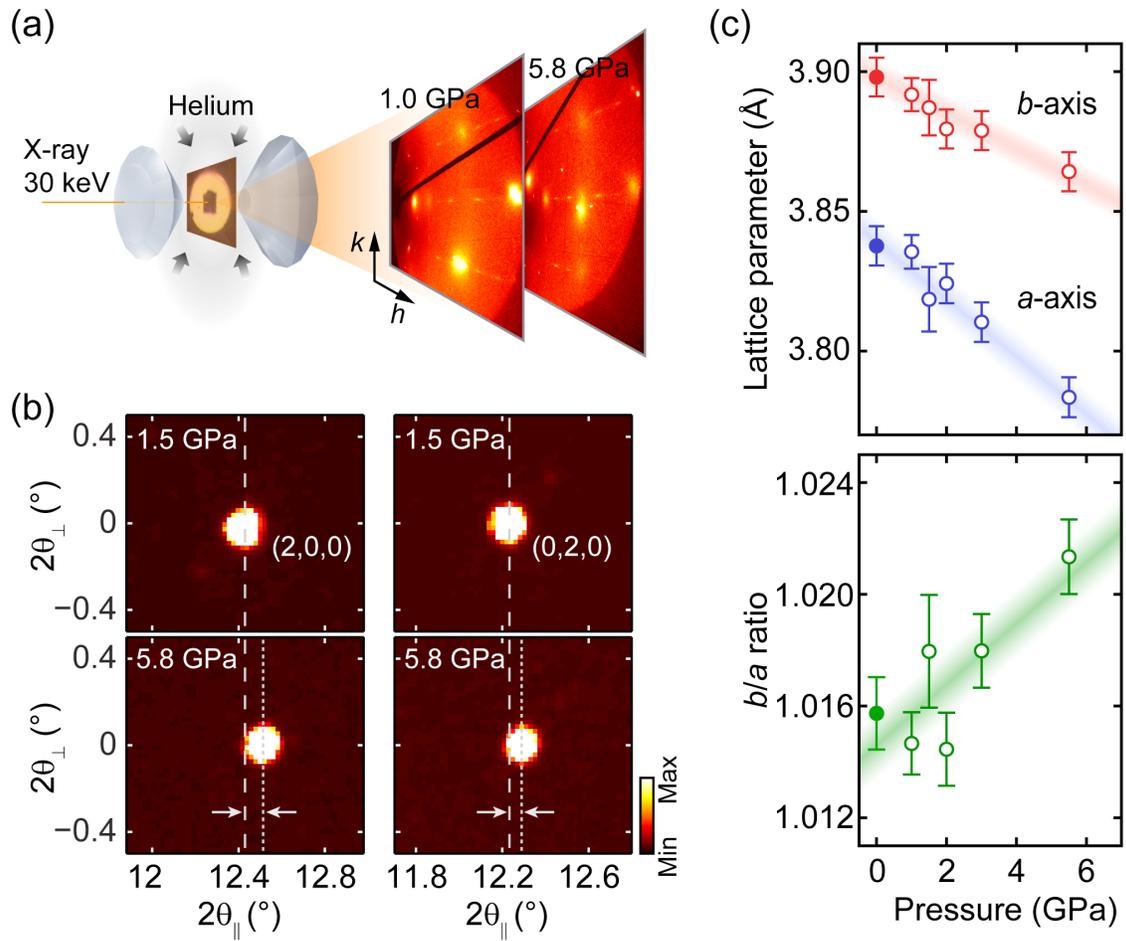

**Fig. 2:** (a) Transmission geometry for x-ray diffraction at hydrostatic high-pressure. Images show 2D diffraction patterns obtained at 1.0 GPa and 5.8 GPa. Black lines on diffraction patterns are the shadows of the beamstop. (b) Zoomed images of lattice peaks – (2, 0, 0) and (0, 2, 0). Dashed and dotted lines denote positions of $2\theta_\parallel$ at 1.5 and 5.8 GPa, respectively. (c) Pressure dependent *a*- and *b*-axis lattice parameters (upper panel) and *b/a* ratio, i.e. orthorhombicity (lower). Note that data at P = 0 (closed circles) are obtained from the RSXS. Error bars correspond to 1 standard deviation (SD).

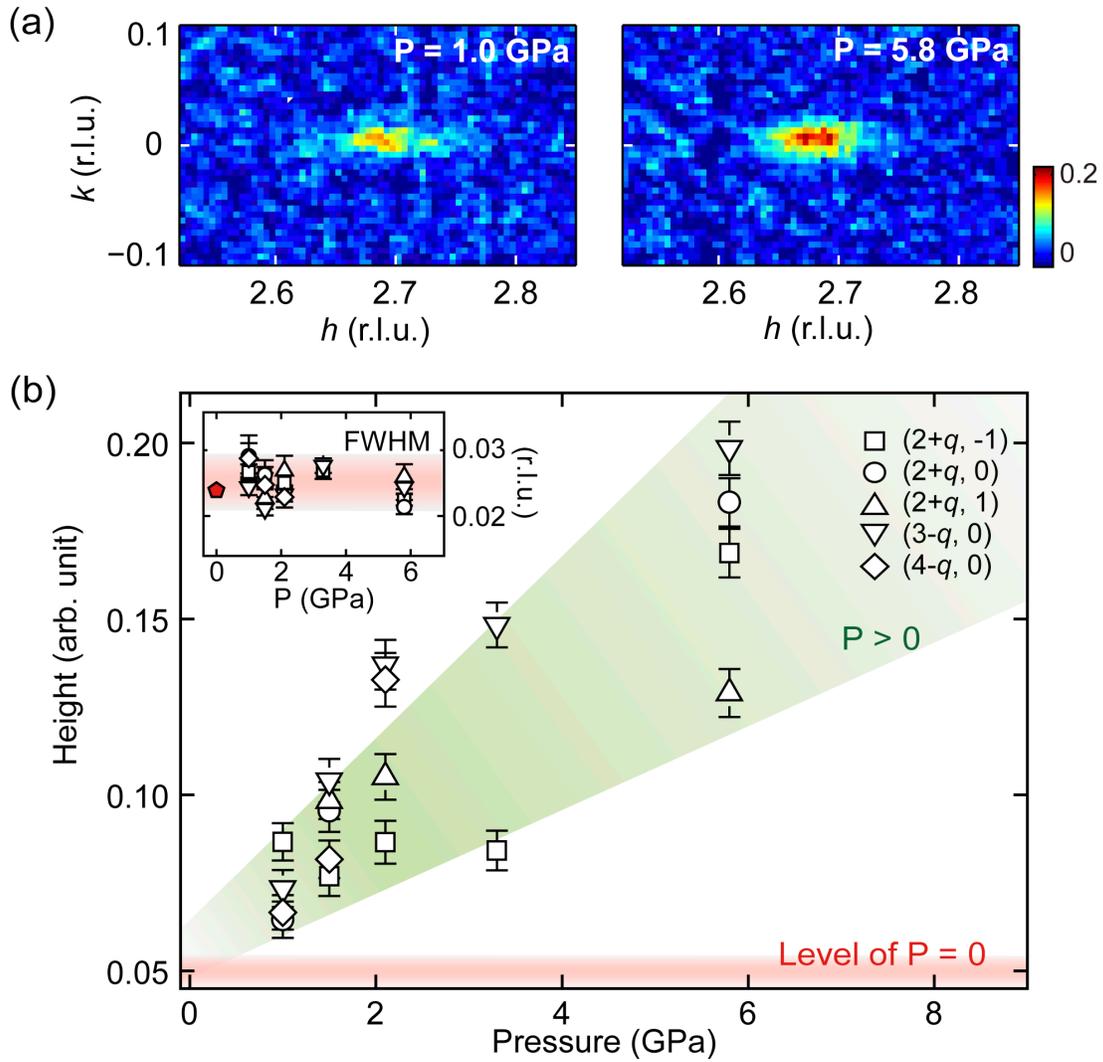

**Fig. 3:** (a) Experimental patterns of OO at $(h, k) = (3\text{-}q, 0)$ with $q = 0.33$ with 1.0 and 5.8 GPa. (b) Summary of fitting results on five OO peaks with varying pressure. Inset shows the full-width half maximum along the $k$-direction. Note that the red colored shading, and data point at P = 0 (closed symbol) are estimated from the RSXS.

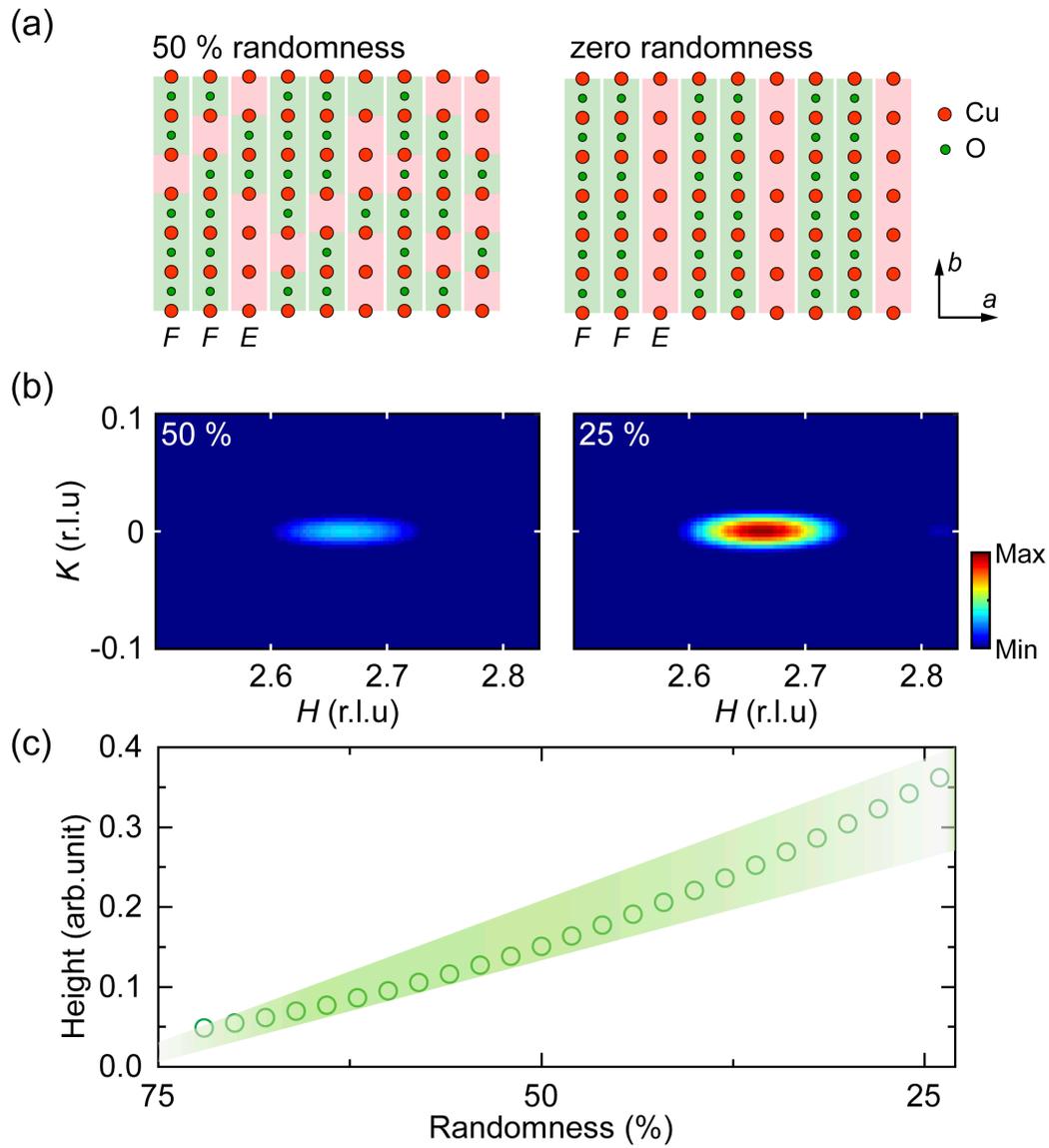

**Fig. 4:** (a) Modeled ortho-III structures with 50 % (left) and 0 % (right) randomness. (b) Simulated OO diffraction patterns of the ortho-III structures with 50 % and 25 % randomness. **(c)** Summary of the height of simulated OO peaks as a function of randomness.

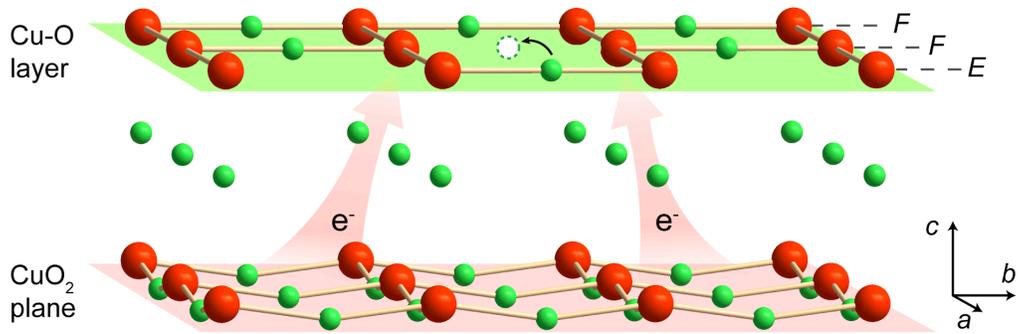

**Fig. 5:** Schematic drawing of charge transfer between Cu-O chain layer and $CuO_2$ plane upon HP. Curved black arrow represents the oxygen movement. The semitransparent arrows schematically represent the charge (electron) transfer from $CuO_2$ plane to Cu-O layer (i.e., hole-doping of $CuO_2$ planes).

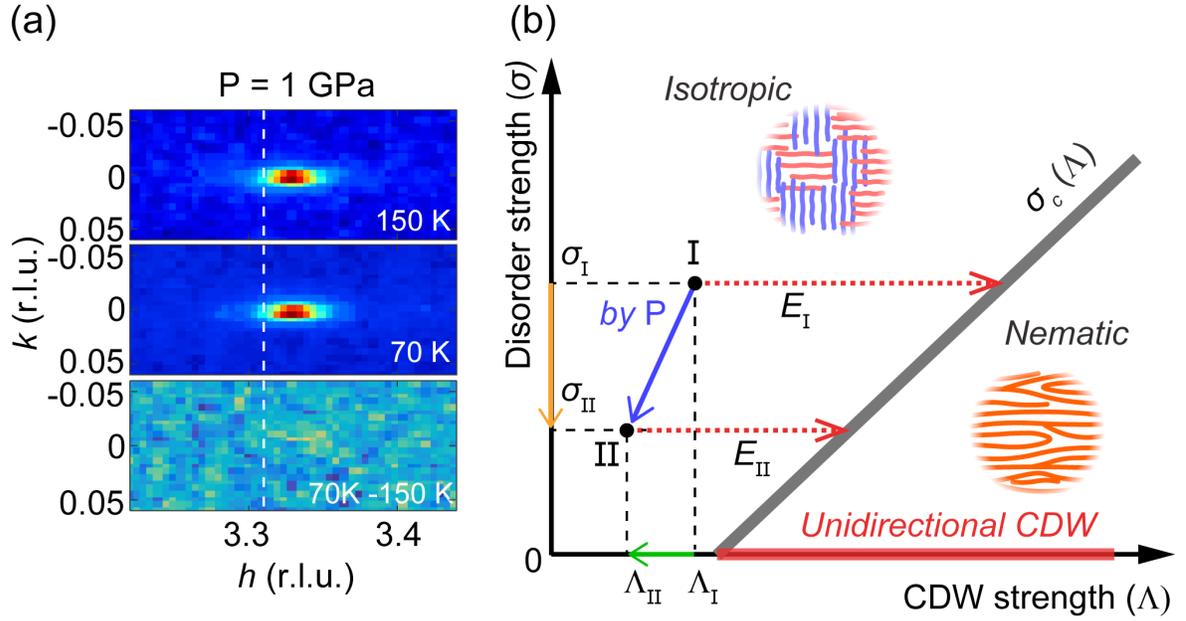

**Fig. 6:** (a) *hk*-diffraction pattern by 30 keV x-ray under P = 1.0 GPa. The bottom panel is extracted by subtracting 150 K data (top) from 70 K data (middle). White dashed-lines denote the CDW *q*-position. (b) Schematic CDW phase diagram (more details in Ref. [39]). The green arrow along the CDW strength ($\Lambda$) axis and the orange arrow along the disorder strength ($\sigma$) axis indicate the directions of pressure effect. Blue and red (dotted) arrows denote total-pressure and magnetic field effects, respectively. Note that state-*I* and state-*II* denote the ambient and compressed YBCO state, respectively.